\documentclass[twoside,twocolumn,english,9pt]{revtex4-1}
\usepackage[T1]{fontenc}
\usepackage[latin9]{inputenc}
\usepackage[a4paper]{geometry}
\usepackage{textcomp}
\usepackage{amsmath}
\usepackage{amssymb}
\usepackage{graphicx}
\usepackage{esint}

\makeatletter
\usepackage{braket}

\usepackage{babel}

\usepackage{babel}

\usepackage{babel}

\usepackage{babel}

\makeatother

\usepackage{babel}
\begin{document}

\title{Fermion confinement via Quantum Walks in 2D+1 and 3D+1 spacetime}

\author{I. Márquez-Martín$^{1}$}

\author{G. Di Molfetta$^{1,2}$}

\email{giuseppe.dimolfetta@lif.univ-mrs.fr}

\author{A. Pérez$\text{\textonesuperior}$}

\affiliation{$^{1}$Departamento de Física Teórica and IFIC, Universidad de Valencia-CSIC,
Dr. Moliner 50, 46100-Burjassot, Spain\\
$^{2}$Aix-Marseille Université, {\'E}cole Centrale de Marseille, Laboratoire d'Informatique
Fondamentale, Marseille, France}

\begin{abstract}
We analyze the properties of a two and three dimensional quantum walk that are
inspired by the idea of a brane-world model put forward by Rubakov
and Shaposhnikov \cite{Rubakov1983}. In that model, particles are
dynamically confined on the brane due to the interaction with a scalar
field. We translated this model into an alternate quantum walk with
a coin that depends on the external field, with a dependence which
mimics a domain wall solution. As in the original model, fermions
(in our case, the walker), become localized in one of the dimensions,
not from the action of a random noise on the lattice (as in the case
of Anderson localization), but from a regular dependence in space.
On the other hand, the resulting quantum walk can move freely along
the ``ordinary'' dimensions. 
\end{abstract}
\maketitle

\section{Introduction}

The quantum walk (QW) is the quantum analogue of the classical random
walk. As in the case of random walks, QWs can appear either under
its discrete-time \cite{Y.AharonovL.Davidovich1993} or continuous-time
\cite{PhysRevA.58.915} form. We will concentrate here on discrete-time
QWs, first considered by Grössing and Zeilinger \cite{grossing1988quantum}
in 1988, as simple one-particle quantum cellular automata, and later
popularized in the physics community in 1993, by Y. Aharonov \cite{Y.AharonovL.Davidovich1993}.
The dynamics of such QWs consists on a quantum particle taking steps
on a lattice conditioned on its internal state, typically a (pseudo)
spin one half system. The particle dynamically explores a large Hilbert
space associated with the positions of the lattice and allows thus
to simulate a wide range of transport phenomena \cite{Kempe2003}.
With QWs, the transport is driven by an external discrete unitary
operation, which sets it apart from other lattice quantum simulation
concepts where transport typically rests on tunneling between adjacent
sites \cite{bloch2012quantum}: all dynamic processes are discrete
in space and time. It has been shown that any quantum algorithm can
be recast under the form of a QW on a certain graph: QWs can be used
for universal quantum computation, this being provable for both the
continuous \cite{PhysRevLett.102.180501} and the discrete version
\cite{PhysRevA.81.042330}. As models of coherent quantum transport,
they are interesting both for fundamental quantum physics and for
applications. An important field of applications is quantum algorithmic
\cite{ambainis2003quantum}. QWs were first conceived as a natural
tool to explore graphs, for example for efficient data searching (see
e.g. \cite{magniez2011search}). They are also useful in condensed
matter applications and topological phases \cite{kitagawa2012observation}.
A totally new emergent point of view concerning QWs concerns quantum
simulation of gauge fields and high-energy physical laws \cite{PhysRevA.94.012335,genske2013electric,Molfetta2016}.
It is important to note that QWs can be realized experimentally with
a wide range of physical objects and setups, for example as transport
of photons in optical networks or optical fibers \cite{schreiber20122d},
or atoms in optical lattices \cite{cote2006quantum}.

Within the context of diffusion processes in lattices, spatial localization
appears as a natural phenomenon. It can result from random noise on
the lattice sites, giving rise to Anderson localization \cite{Lattices1956},
but it can also be driven by the action of an external periodic potential
(see e.g. \cite{aubry1980analyticity,PhysRevLett.49.833,PhysRevLett.103.013901}).
Similarly, one obtains localization for the 1-dimensional QW when
spatial disorder is included \cite{Joye2010,PhysRevLett.106.180403,Crespi2013a},
via non-linear effects \cite{Navarrete-Benlloch2007}, or using a
spatially periodic coin \cite{PhysRevE.82.031122}. For higher dimensions,
localization may appear, even in the noiseless case, from the choice
of the coin operator \cite{PhysRevA.69.052323}.

In this paper, we will propose a different variant of the QW that
gives rise to localization, by introducing a site-dependent non-periodic
coin operator. The model is inspired on a brane-world proposal with
extra dimensions \cite{Rubakov1983}, where particles are confined
to live in the ordinary 3+1 dimensions by the action of a potential
well created by some additional scalar field. In its simplest form,
one accounts for massless fermions which are confined in the brane.
This idea can be translated to describe a QW where the potential well
manifests as a position-dependent coin operator. Differently to the
situations described above, the confining field is not random nor periodic,
being instead a monotonous function of the position. As we show, this
kind of QW produces a dynamical localization of the QW as in the original
model. In fact, it can be shown that, in the continuous space-time
limit, one reproduces the dynamics of a massless Dirac fermion. In
this way, we establish an interesting parallelism between a high-energy
quantum field theory, and a QW model that results in localization.

The rest of this paper is organized as follows. In Sect. II we briefly
introduce the original brane model \cite{Rubakov1983} that motivated
our work. In Sect. III we make use of this model to introduce a QW
on two dimensions with a position-dependent coin that simulates the
domain wall ``scalar field'' along the second (or ``extra dimension'').
We show that this QW in fact results in a confinement of the walker,
and that the space-time continuous limit indeed reproduces the dynamics
of a Dirac particle coupled to the scalar field. These ideas are generalised to 3D in Sect. IV.  
Finally, Sect. V is devoted to summarizing and discussing our results.

\section{Domain wall model for particle physics}

\label{sec:sec1}

The possibility of extra dimensions of space was first suggested by
Theodor Kaluza and Oscar Klein \cite{Kaluza1921,Klein1926} seeking
for an unified theory of electromagnetic and gravitational fields
into a higher dimensional field, with one of the dimensions compactified.
However, experimental data from particle colliders restrict the compactification
radius to such small scales that they become virtually impossible
to access them experimentally. A way to overcome this difficulty \cite{Arkani-Hamed1999}
makes use of the ideas put forward by Rubakov and Shaposhnikov \cite{Rubakov1983}.
In that paper, the authors propose a brane world scenario, in which
space-time has (3+$N$)+1 dimensions, with ordinary (low energy) particles
confined in a potential well which is narrow along $N$ spatial directions
and flat along the remaining three directions. The origin of this
potential well is suggested to have a dynamical origin. In the simplest
case it can be created by an extra scalar field in $4+1$ dimensions,
as described by the Lagrangian 
\begin{equation}
\mathcal{L}=\frac{1}{2}\partial_{A}\partial^{A}\varphi-\frac{1}{2}m^{2}\varphi-\frac{1}{4}\lambda\varphi^{4},\,\,\,A=0,1,2,3,4,\label{eq:Lagrangianscalar}
\end{equation}
with metrics $g_{AB}=(1,-1,-1,-1,-1)$. The classical equations of
motion derived from the above Lagrangian admit a domain wall solution
$\varphi(x^{4})$ that only depends on the coordinate $x^{4}$ along
the extra dimension, and is given by 
\begin{equation}
\varphi(x^{4})=\frac{m}{\sqrt{\lambda}}\tanh(\frac{mx^{4}}{\sqrt{2}}).\label{eq:phidomainwall}
\end{equation}
This model can account for left-handed massless fermions living in
$3+1$ dimensions, if they are coupled to the scalar fields, as in
the following Lagrangian: 
\begin{equation}
\mathcal{L}_{\psi}=i\bar{\Psi}\Gamma^{A}\partial_{A}\Psi+h\varphi\bar{\Psi}\Psi,\label{eq:Lagrangianfermion}
\end{equation}
where $h$ is the coupling constant, and the $4+1$-dimensional gamma
matrices are $\Gamma^{\mu}=\gamma^{\mu}$, $\mu=0,\dots3$, and $\Gamma^{4}=i\gamma^{5}$,
with $\gamma^{\mu},\gamma^{5}$ the standard gamma matrices. From
Eq. (\ref{eq:Lagrangianfermion}) the corresponding Dirac equation
follows, which reads 
\begin{equation}
i\Gamma^{A}\partial_{A}\Psi+h\varphi\Psi=0.\label{eq:Diracfermion}
\end{equation}
As discussed in \cite{Rubakov1983}, this equation has a solution
that is confined inside the domain wall, while the corresponding particles
are left-handed massless fermions in the $3+1$ dimensional world.
In the next Section, we make use of these ideas to introduce a QW model
in $1+1+1$ dimensions that leads to confined fermions in $1+1$.

\section{2D Quantum Walks inside a 1+1 Domain Wall}

\label{sec:sec2}

Consider a QW defined over discrete time and discrete two-dimensional
space, with axis $x$, $y$. The discrete space points are labeled
by $p$ and $q$, respectively, with $p,q\in\mathbb{Z}$, while time
steps are labeled by $j\in\mathbb{N}$. This QW is driven by an in-homogeneous
coin acting on the $2$-dimensional Hilbert space $\mathcal{H}_{\text{spin}}$.
The evolution equations read

\begin{equation}
\begin{bmatrix}\psi_{j+1,p,q}^{\uparrow}\\
\psi_{j+1,p,q}^{\downarrow}
\end{bmatrix}\ =S_{y}Q^{+}(\theta_{q})S_{x}Q^{-}(\theta_{q})\begin{bmatrix}\psi_{j,p,q}^{\uparrow}\\
\psi_{j,p,q}^{\downarrow}
\end{bmatrix},\label{eq:defwalkdiscr}
\end{equation}

with $Q^{\pm}(\theta_{q})$ defined as

\begin{equation}
Q^{\pm}(\theta_{q})=\begin{pmatrix}\cos\theta_{q}^{\pm} & i\sin\theta_{q}^{\pm}\\
i\sin\theta_{q}^{\pm} & \cos\theta_{q}^{\pm}
\end{pmatrix},
\end{equation}

where $\theta_{q}^{\pm}$ = $\pm\frac{\pi}{4}-\epsilon\bar{\theta}_{q}$
is the coin angle, which depends only on the coordinate $q$, and
$\epsilon$ is a small parameter that allows to reach the appropriate
continuous space-time limit (see discussion below). The operators
$S^{x}$ and $S^{y}$ are the usual spin-dependent translations along
the x-direction and the y-direction, respectively. They are defined
as follows: 
\begin{equation}
S^{x}\Psi_{j,p,q}=\left(\psi_{j,p+1,q}^{\uparrow},\psi_{j,p-1,q}^{\downarrow}\right)^{\top},
\end{equation}
and 
\begin{equation}
S^{y}\Psi_{j,p,q}=\left(\psi_{j,p,q+1}^{\uparrow},\psi_{j,p,q-1}^{\downarrow}\right)^{\top}.
\end{equation}

\begin{figure}
\includegraphics[width=1\columnwidth]{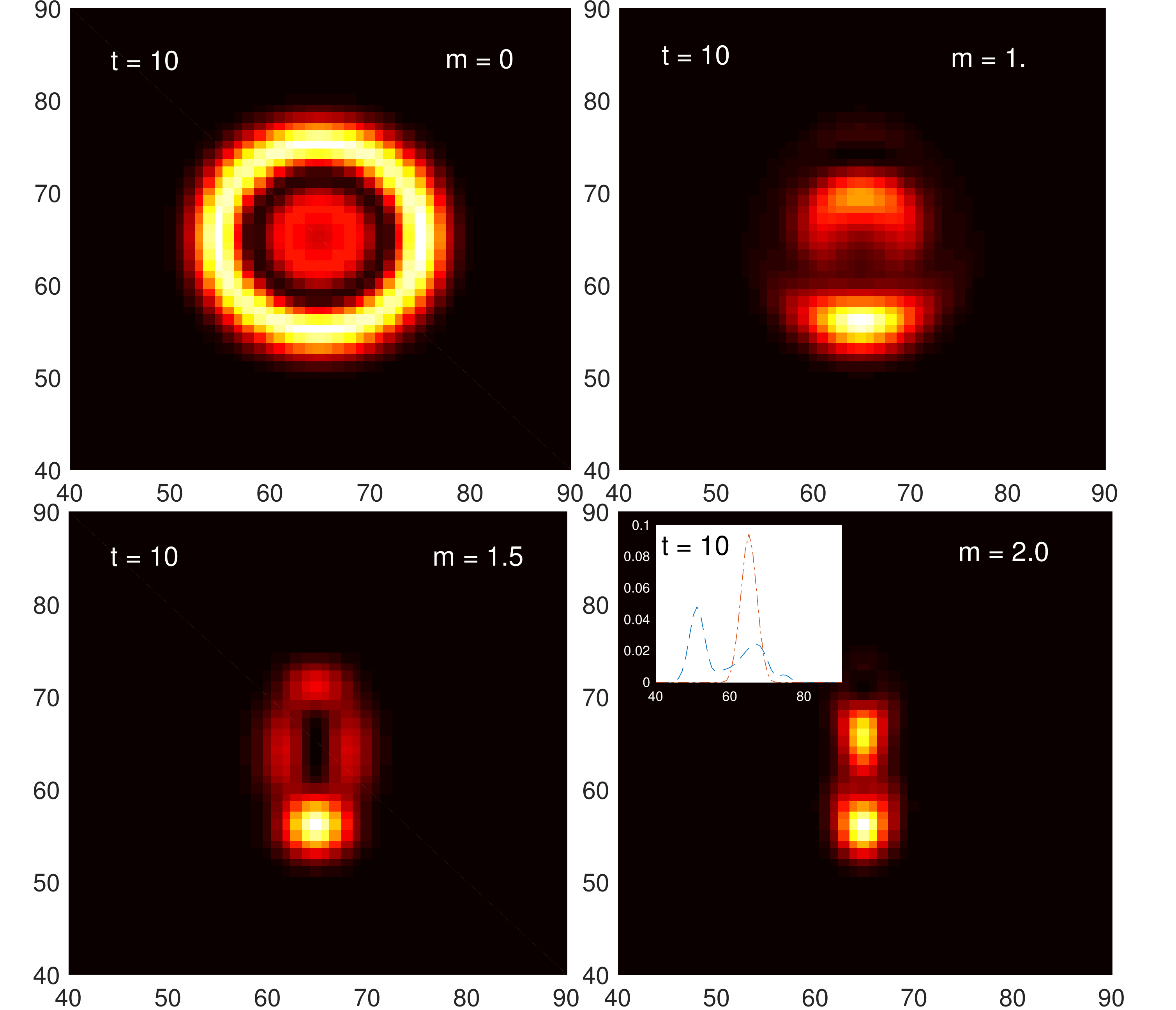}

\caption{(Color online) Probability distribution $||\Psi(t_{j},x_{p},y_{q})||^{2}$
of the two-dimensional QW for a value $t=10$ of the timestep, and
different values of $m$. The rest of parameters are fixed to $\lambda=60$,
$h=70$, with the lattice parameter $\epsilon=0.04$. The inset in
the last subfigure also shows the projected density profile along each direction
of the lattice (\textit{red dot-dashed line} represents the x-direction
and \textit{blue dashed line} the y-direction). The initial condition
is a Gaussian wave packet $\Psi(0,x_{p},y_{q})=\sqrt{n(x_{p},y_{q})}\otimes(\frac{1}{\sqrt{2}},\frac{1}{\sqrt{2}})^{\top}$
centered at the point $(64,64)$, where the gaussian distribution
$n(x_{p},y_{q})$ has a width $\delta=0.1$. }
\label{fig:Probability-density-of-2} 
\end{figure}

\begin{figure}
\includegraphics[width=1\columnwidth]{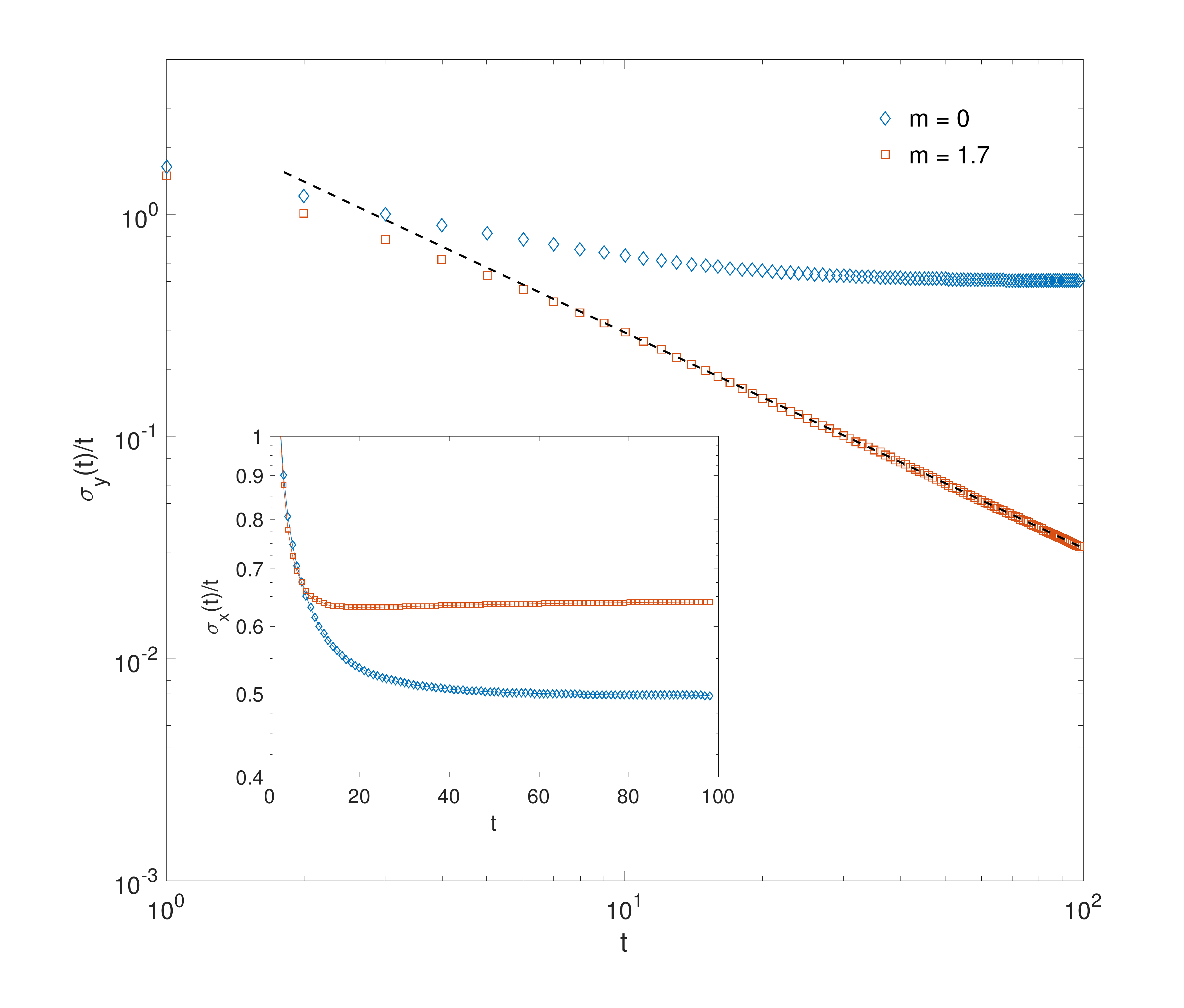} \caption{(Color online) Time evolution of the standard deviation divided by
the timestep, i.e., $\sigma_{x}(t)/t$ (in the inset) and $\sigma_{y}(t)/t$,
calculated independently along the $x$ and $y$ directions, for a
localised (red squares) and a free fermion (blue diamonds). The initial
condition is a Gaussian wave packet $\Psi(0,x_{p},y_{q})=\sqrt{n(x_{p},y_{q})}\otimes(0,1)^{\top}$
centered around $(128,128)$, and the parameters of the potential
are $\lambda=60$ and $h=70$, with the lattice parameter $\epsilon=0.02$.}
\label{fig:sigma} 
\end{figure}

Equations (\ref{eq:defwalkdiscr}) describe the evolution of a two-level
system, e.g., a fermion in two dimensions, and it has been shown that
each of them recover, in the continuous limit, the Dirac equation
\cite{di2014quantum}, where the parameter $\theta_{q}$ corresponds
to a position-dependent potential. Let us now consider $\bar{\theta}_{q}$
of the form: 
\begin{equation}
\bar{\theta}_{q}=h\frac{m}{\sqrt{\lambda}}\tanh(\frac{mq}{\sqrt{2}}),
\label{eq:potential}
\end{equation}
and notice that it corresponds to a narrow potential in the $q$-direction
when $m$, the \textquotedbl{}effective mass\textquotedbl{} is sufficiently
large.

Fig. \ref{fig:Probability-density-of-2} shows the evolved probability
distribution of this 2D QW, starting from a symmetric Gaussian profile
in both directions. As the mass is increased, the probability becomes
strongly localized around the $y-$axis, while it evolves as a usual QW
on the non-confining $x-$direction. This features are clearly seen
in Fig. \ref{fig:sigma}, where we have represented the standard deviation
divided by the timestep, i.e., $\sigma_{x}(t)/t$ and $\sigma_{y}(t)/t$,
calculated independently along the $x$ and $y$ directions. For $m=0$
(no confinement), both quotients tend to a constant, which corresponds
to the normal spreading of a 2D QW in both directions. As $m$ increases,
localization acts on the $y$- direction, and manifests as an exponential
decay of $\sigma_{y}(t)/t$. On the other hand, the standard deviation
corresponding to the $x$ axis behaves as a free-evolving QW, with
a spreading velocity that depends on the parameters of the potential
well.

As we show below, in the continuous limit equations (\ref{eq:defwalkdiscr})
are in correspondence with Eq. (\ref{eq:Diracfermion}), describing
the propagation of a massless fermion in a space-time manifold $M^{(1+N,1)}$,
the usual Minkowski space with $1+N$ spatial dimensions. When $m$
is non-vanishing, the fermion is confined inside a potential well,
which is sufficiently narrow along $N$ directions and flat along
the other one (in our case $N=1$).

Let us introduce new space-time coordinates $t_{j}$, $x_{p}$ and
$y_{q}$ such that $t_{j}=j\epsilon$, $x_{p}=p\epsilon$ and $y_{q}=q\epsilon$.
In the limit when $\varepsilon\longrightarrow0$, these coordinates
become continuous, labeled by $t$, $x$ and $y$, respectively. If
we Taylor expand equations (\ref{eq:defwalkdiscr}) around $\epsilon=0$,
we recover the following equation: 
\begin{equation}
\partial_{t}\Psi(t,x,y)=\left[\sigma_{z}\partial_{x}-\sigma_{y}\partial_{y}-i\sigma_{x}\bar{\theta}(y)\right]\Psi(t,x,y),
\end{equation}
which can be recast in covariant form: 
\begin{equation}
i\Gamma^{A}\partial_{A}\Psi+h\frac{m}{\sqrt{\lambda}}\tanh(\frac{my}{\sqrt{2}})\Psi=0,\label{eq:DiracKK}
\end{equation}
where $\Gamma^{A}=\{\gamma^{\mu},\gamma^{c}\}$, $\mu=0,1$ and $\gamma^{c}=i\gamma^{5}=i\gamma^{0}\gamma^{1}=-i\sigma_{z}$.
In this equation, $\gamma^{0}=-\sigma_{x}$, $\gamma^{1}=-i\sigma_{y}$.
As can be easily seen, Eq. (\ref{eq:DiracKK}) takes the same form
as (\ref{eq:Diracfermion}) if we make the identification $x^{4}\longrightarrow y$
and $\varphi\longrightarrow\frac{m}{\sqrt{\lambda}}\tanh(\frac{my}{\sqrt{2}})$.

\section{3D Quantum Walks inside a 2+1 Domain Wall}
\label{sec:sec3}

The extension of the previous case to the higher dimensional case is straightforward. 
In this section we adopt the same techniques introduced in the last section but we double the spin Hilbert space, in order to recover the standard Dirac equation in 3+1 spacetime. Let us recall that in 3+1, gamma matrices appearing in equation (\ref{eq:Diracfermion}), are four dimensional. In the Weyl representation they read:
\begin{align}
\gamma^0 = \begin{pmatrix} 0 & \mathbb{I}  \\ \mathbb{I} &  0  \end{pmatrix} \ & \ \gamma^i = \begin{pmatrix} 0 & \sigma^i  \\ -\sigma^i &  0  \end{pmatrix}\ \gamma^5 = \begin{pmatrix} -\mathbb{I} & 0   \\ 0  &  \mathbb{I}  \end{pmatrix}.
\end{align}

\begin{figure}
\includegraphics[width=0.8\columnwidth]{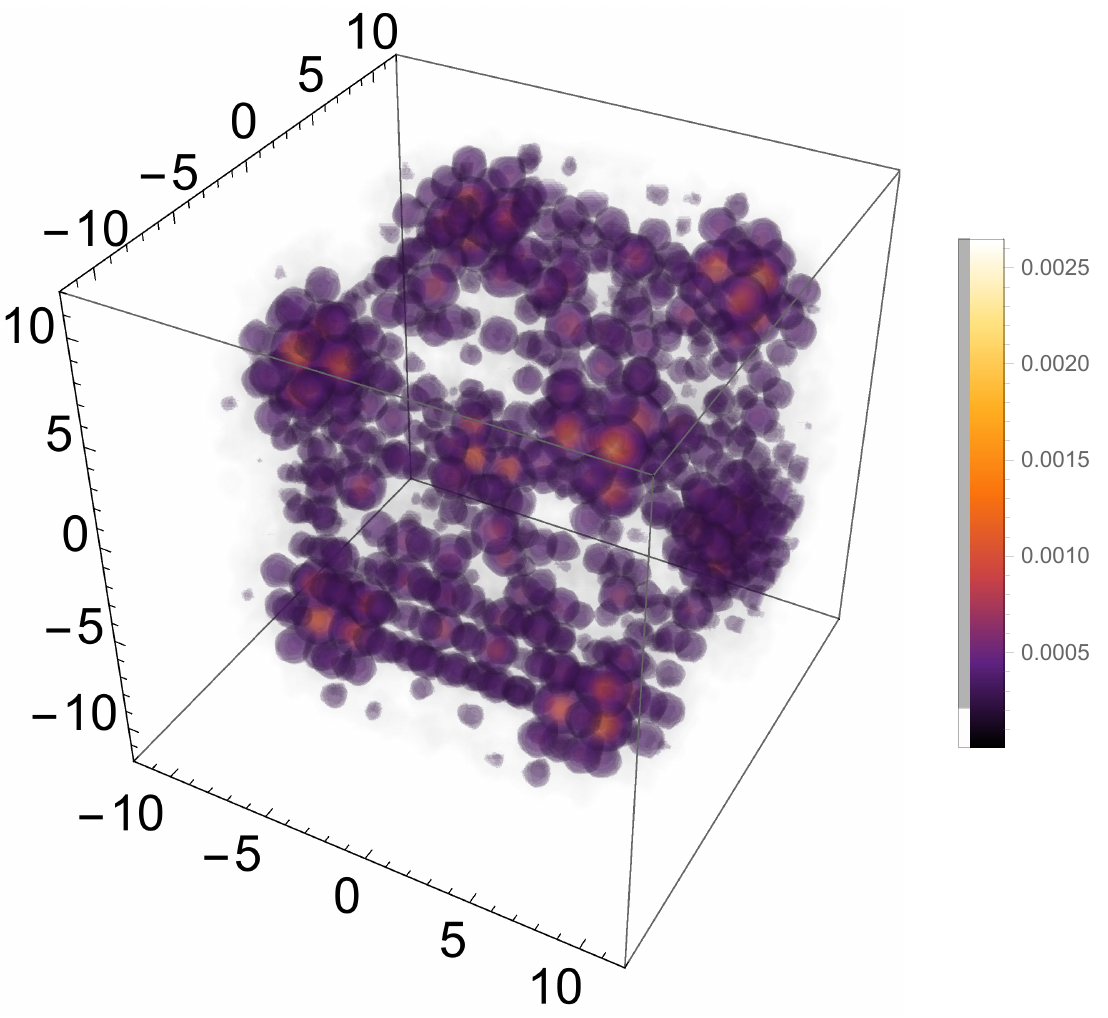}
\caption{(Color online) Density plot in 3D at time $j=12$ with Gaussian initial wave packet $\Psi(0,x_{p},y_{q},z_{r})=\sqrt{n(x_{p},y_{q},z_{r})}\otimes(1,i,1,i)^{\top}$
centered around $(0,0)$, and for $m$ = $0$.} 
\label{fig:3DQW} 
\end{figure}

Now, consider the QW defined over discrete three-dimensional
space, with axis $x$, $y$ and $z$. The discrete space points are labeled by $p$, $q$ and $r$, respectively, with $p,q,r \in\mathbb{Z}$. This QW is driven by an in-homogeneous
coin acting on the spinor $\left( \psi^1_{j,p,q,r},\psi^2_{j,p,q,r} \right)^\top$, where each $\psi^i_{j,p,q,r}$ belongs to $\mathcal{H}_{\text{spin}}$ for $i=1,2$. 

The evolution equations read:

\begin{equation}
\begin{bmatrix}\psi^1_{j+1,p,q,r}\\
\psi^2_{j+1,p,q,r}
\end{bmatrix}\ = \Theta_r \mathcal{S}^{z}\mathcal{R}_z\mathcal{S}^{x}\mathcal{R}_x\mathcal{S}^{y}\mathcal{R}_y
\begin{bmatrix}\psi^1_{j+1,p,q,r}\\
\psi^2_{j+1,p,q,r}
\end{bmatrix},\label{eq:defwalkdiscr3D}
\end{equation}

where
\begin{align}
\Theta_r = \begin{pmatrix} \cos\bar\theta_r \epsilon & i \sin\bar\theta_r \epsilon\\ i \sin\bar\theta_r \epsilon & \cos\bar\theta_r \epsilon\end{pmatrix} \otimes \mathbb{I}_{2},
\end{align}
and 
\begin{align}
\mathcal{S}^{i} =\begin{pmatrix}
S^i & 0 \\  0 & {S^i}^\dagger
\end{pmatrix} & \hspace{1.5cm} \mathcal{R}_{i} =\begin{pmatrix}
R_i & 0 \\  0 & R_i
\end{pmatrix},&
\end{align}

where the operators $S^{i}$ are the usual spin-dependent translations along each direction 
of the cubic lattice, and each unitary rotation $R_i$, for $i = x,y,z$ is an element of $U(2)$.

\begin{figure}
\includegraphics[width=0.7\columnwidth]{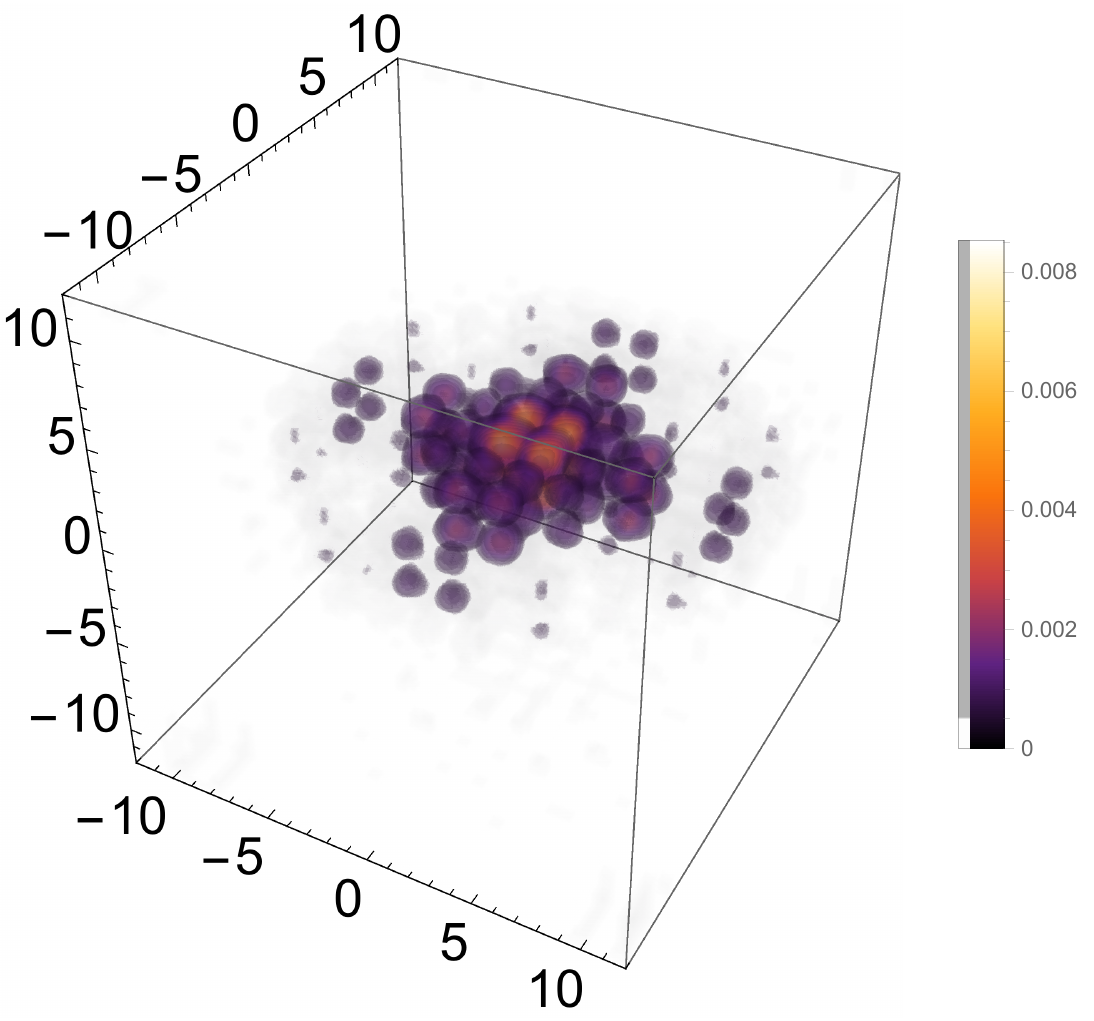}
\includegraphics[width=0.4\columnwidth]{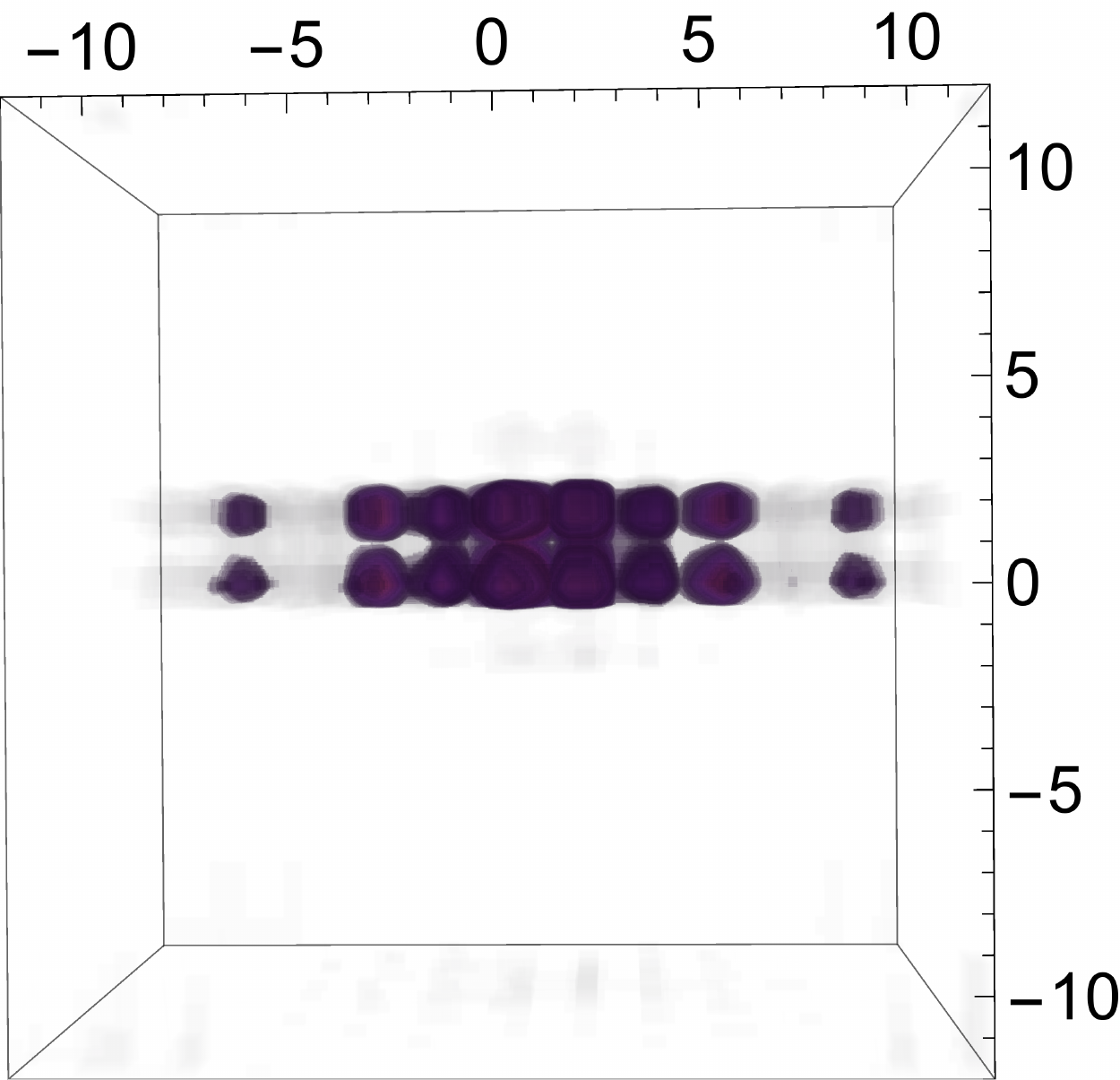}
\includegraphics[width=0.4\columnwidth]{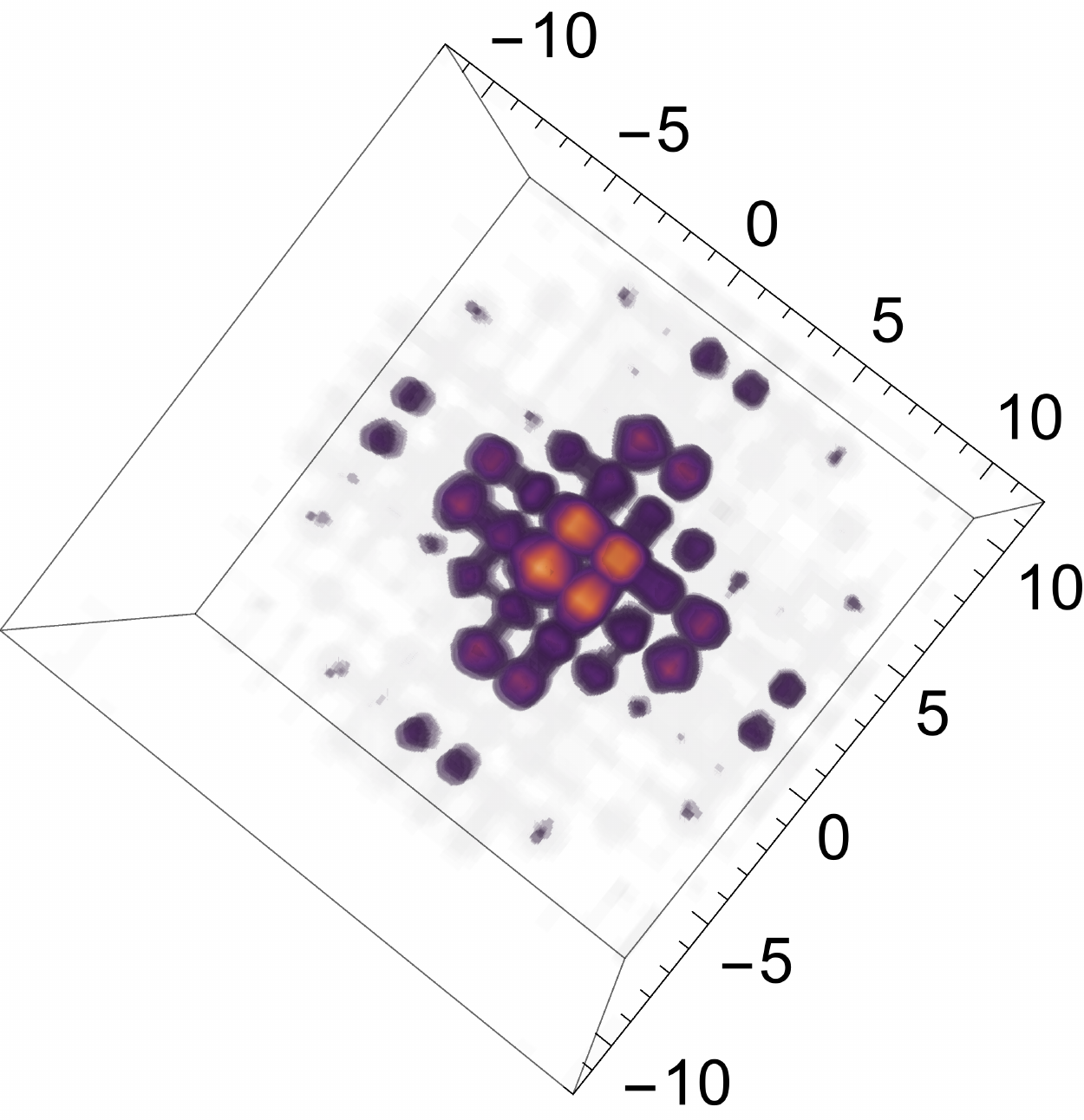}
\caption{(Color online) Density plots in 3D at time $j=20$ with Gaussian initial wave packet $\Psi(0,x_{p},y_{q},z_{r})=\sqrt{n(x_{p},y_{q},z_{r})}\otimes(0,1,0,1)^{\top}$
centered around $(0,0)$. The parameters of the potential are $\lambda=90$, $h=4$ and $m$=$11$. The two subfigures at the bottom display the $x$-$z$ side view (left) and the $x$-$y$ side view (right) of the 3D density plot.} 
\label{fig:brane} 
\end{figure} 

Notice that $\Theta_r$ encodes the coupling between the spinor components, and $\theta_r$ is an arbitrary position-dependent function, which can model either the mass term or any other scalar potential. If $\theta_r$ identically vanishes, Eq. ($\ref{eq:defwalkdiscr3D}$) represents simply a couple of independent split-step QW operators acting on each component of the spinor. In the following, this mass-term is defined by Eq. (\ref{eq:potential}), and will model the narrow potential in the $r$ direction, embedding a 3D QW in a 2D spacetime lattice. 

In order to validate the model, we compute the formal continuous limit of Eq. ($\ref{eq:defwalkdiscr3D}$) with same technique introduced in the previous section. Thus, let us introduce the new spatial coordinate $z_{r}$, such that $z_{r}=r\epsilon$, and again assume that in the limit when $\varepsilon\longrightarrow0$, this coordinate, together with $t_j$, $x_p$, $y_q$, become continuous, labeled by  $z$ and $t$, $x$, $y$, respectively. If
we Taylor expand equations ($\ref{eq:defwalkdiscr3D}$) around $\epsilon=0$, the zero order restricts the four-dimensional coins, $\mathcal{R}_i = R_i \otimes \mathbb{I}_{2}$:
\begin{equation}
\begin{bmatrix}\psi^1\\
\psi^2
\end{bmatrix}\ =  \mathcal{R}_z\mathcal{R}_x\mathcal{R}_y
\begin{bmatrix}\psi^1\\
\psi^2
\end{bmatrix} + O(\epsilon),\label{eq:CL}
\end{equation}
which leads to the condition
\begin{equation}
\mathcal{R}_z \mathcal{R}_x \mathcal{R}_y = \mathbb{I}_{4}.
\end{equation}

Then the first order term of the Taylor expansion reads:

\begin{equation}
\partial_t\begin{bmatrix}\psi^1\\
\psi^2
\end{bmatrix}\ = (B_z \partial_z + B_x \partial_x + B_y \partial_y + i B_0 \bar\theta(z))
\begin{bmatrix}\psi^1\\
\psi^2
\end{bmatrix} + O(\epsilon),\label{eq:CL3D}
\end{equation}

where
\begin{align}
B_z = Z \mathcal{R}_z\mathcal{R}_x\mathcal{R}_y\nonumber\\ 
B_x = \mathcal{R}_z Z \mathcal{R}_x\mathcal{R}_y\nonumber\\
B_y = \mathcal{R}_z \mathcal{R}_x Z \mathcal{R}_y,
\end{align}
and 
\begin{align}
B_0 = \sigma_x \otimes \mathbb{I}_2\nonumber\\
Z = \mathbb{I}_2 \otimes \sigma_z.
\end{align}
Now, comparing Eq. (\ref{eq:CL3D}) with equation (\ref{eq:Diracfermion}) we derive - up to a U(2) rotation - the explicit form of each rotation $\mathcal{R}_i$. In particular, we need to satisfy $\gamma^0\gamma^1 = B_x$, $\gamma^0\gamma^2 = B_y$ and $\gamma^0\gamma^3 = B_z$, which leads to:
\begin{align}
R_x  = \frac{1}{\sqrt{2}} \begin{pmatrix}1 & 1 \\ 1 & -1 \end{pmatrix} \hspace{1.5cm}  R_z  = \frac{1}{\sqrt{2}} \begin{pmatrix}1 & -i \\ i & -1 \end{pmatrix}  \hspace{0.25cm}   \nonumber \\
R_y = R_x R_z.\hspace{3cm}
\end{align}

Thus, numerical simulations of the above QW can model the behavior of a fermion in a 3+1 space time. In particular, in Fig. \ref{fig:3DQW}, the quantum walker spreads on the 3D cubic lattice, starting from a symmetric initial condition, recovering in the continuous limit, a massless fermion in vacuum ($\bar\theta$ = $0$). 
In contrast, Fig. \ref{fig:brane} shows the evolved probability distribution of this 3D QW when the mass-term is different from zero and position-dependent. As in the lower dimensional case, the probability dynamically localises on the $x$-$y$ plane, and corresponds to a standard 2D QW, while it possesses a finite size on the $z$-direction, which typically decreases with the lattice parameter $\epsilon$. \\

\section{Discussion}

\label{sec:Discussion}

In this paper we have studied the properties of a two and a three dimensional 
QW that are inspired by the idea of a brane-world model put forward
by Rubakov and Shaposhnikov \cite{Rubakov1983}. In that model, particles
are dynamically confined in the brane due to the interaction with
a scalar field. We translated this model into an alternate QW with
a coin that depends on the external field, with a dependence which
mimics a domain wall solution. As in the original model, fermions
(in our case, the walker), become confined in one of the dimensions,
while they can move freely on the ``ordinary'' dimensions. In this
way, we can think of the QW as a possibility to simulate brane models
of quantum field theories. In the opposite direction of thought, we
obtain a QW that shows localization, not from random noise on the
lattice or from a periodic coin, as in previous models, but from a
coin which changes in space in a regular, non periodic, manner. In
our opinion, this interplay between QWs and high energy theories can
be beneficial for both fields.

\section{Acknowledgements}

This work has been supported by the Spanish Ministerio de Educación
e Innovación, MICIN-FEDER project FPA2014-54459-P, SEV-2014-0398 and
Generalitat Valenciana grant GVPROMETEOII2014-087.

 \bibliographystyle{apsrev4-1}
\bibliography{library}

\end{document}